\documentclass[range,oneside,]{img/ar2e}
\usepackage{img/cite,img/psfig,img/ulem}
\usepackage{amsmath,amssymb,siunitx,graphicx,amsfonts,bm,array,dcolumn}

\usepackage{xcolor,hyperref}
\definecolor{blue}{rgb}{0.,0.,0.5}   
\hypersetup{
	colorlinks,
	linkcolor={blue},
	citecolor={blue},
	urlcolor={blue}
}
\definecolor{orange}{rgb}{1,0.3,0}   
\usepackage[sort&compress,numbers]{natbib}
\usepackage{doi}
\usepackage{setspace}
\usepackage{longtable}
\usepackage{array}
\usepackage{dcolumn}
\usepackage{booktabs}
\usepackage{multirow}
\usepackage{graphicx}

\newcolumntype{d}{D{.}{.}{-1}} 

\newcommand{\bra}[1]{\ensuremath{\langle #1|}}	
\newcommand{\ket}[1]{\ensuremath{|#1\rangle}}	
\newcommand{\braket}[1]{\ensuremath{\langle #1\rangle}}	
\renewcommand{\v}[1]{\ensuremath{\boldsymbol{#1}}}		
\newcommand{\E}[1]{\ensuremath{\times10^{#1}}}	

\newcommand{\g}{\ensuremath{\gamma}} 
\renewcommand{\a}{\ensuremath{\alpha}} 
\newcommand{\s}{\ensuremath{\sigma}} 
\newcommand{\Q}{\ensuremath{Q_{W}}}
\newcommand{\+}{\ensuremath{^{+}}}		
\newcommand{\el}[1]{\ensuremath{{}^{#1}}}	
\renewcommand{\d}{\ensuremath{{\rm d}}}	

\renewcommand{\P}{\ensuremath{{P}}}
\newcommand{\C}{\ensuremath{{C}}}
\newcommand{\T}{\ensuremath{{T}}}

\newcommand{\wf}{wavefunction}
 
\renewcommand{\contentsname}{\\[-1.3cm]} 

\newcommand{\sidebar}[2]{\marginpar{{\bf #1}---\footnotesize{#2}}}

\newcommand{\PBD}{Porsev2009,Porsev2010}
\newcommand{\QED}{Kuchiev2002a,Milstein2003,Johnson2001,Sapirstein2003a}
\newcommand{\RobertsCosmic}{RobertsCosmic2014,RobertsCosPRD2014}
\newcommand{\kvi}{NunezPortela2013} 
\newcommand{\triumf}{Tandecki2014} 

\begin{document}
\title{Parity and Time-Reversal Violation in Atomic Systems}

\author{
	~~~~~~~~~~B. M. Roberts, V. A. Dzuba, and V. V. Flambaum
	\affiliation{School of Physics, University of New South Wales, Sydney, NSW 2052, Australia}%
\\[-2cm]{\begin{center}\textnormal{\small(Dated: \today)}\end{center}}\\[-0.75cm]   
}%
\markboth{{Roberts, Dzuba \& Flambaum}}{Parity and Time-Reversal Violation in Atomic Systems}
\begin{abstract} 
Studying the violation of parity and time-reversal invariance in atomic systems has proven to be a very effective means for testing the electroweak theory at low energy and searching for physics beyond it. 
Recent developments in both atomic theory and experimental methods have led to the ability to make extremely precise theoretical calculations and experimental measurements of these effects.
Such studies are complementary to direct high-energy searches, and can be performed for just a fraction of the cost.
We review the recent progress in the field of parity and time-reversal  violation in atoms, molecules, and nuclei, and examine the implications for physics beyond the Standard Model, with an emphasis on possible areas for development in the near future.\\[-0.0cm]

{ This is a preprint of a manuscript prepared for publication in {\sl Annual Review of Nuclear and Particle Science}, vol.~{\bf65} (2015).}   
\end{abstract}
\maketitle

\section{INTRODUCTION}\label{sec:intro}

\begin{figure}[b!]
\centering
\includegraphics[width=0.25\textwidth]{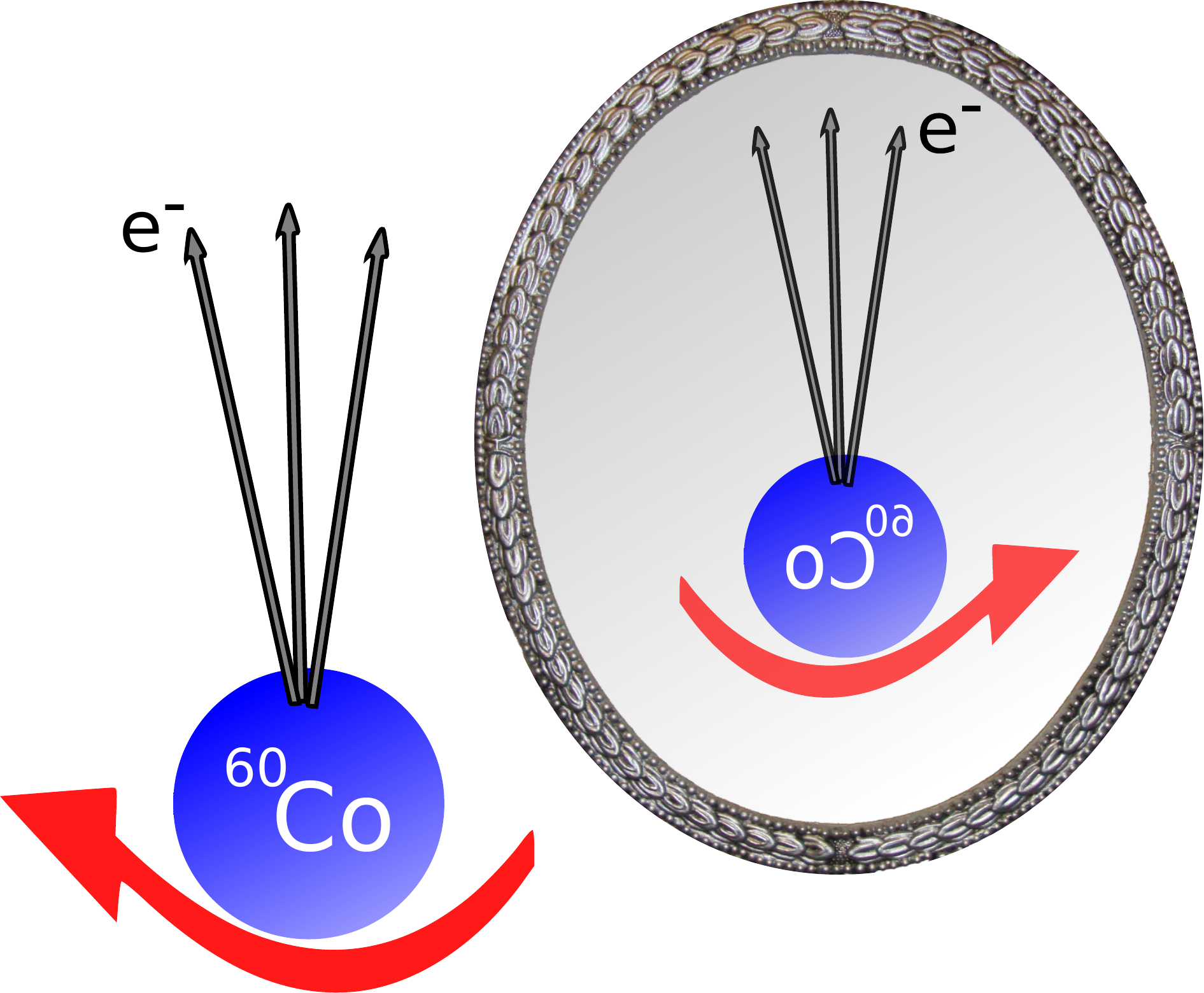}
\caption{\small
	The distribution of electrons emitted in the $\beta$-decay of polarized \el{60}Co nuclei was observed to be anisotropic, 
providing unequivocal proof of parity violation \cite{Wu1957}.
}
\label{img:co}
\end{figure}

Parity violation was first observed by Wu {\sl et al.}~\cite{Wu1957} in 1957, not long after Lee and Yang made their Nobel prize winning suggestion that parity may not be conserved in weak interactions \cite{LeeYang1956}; see Fig.~\ref{img:co}.
Atomic parity nonconservation (PNC) is caused by the weak interaction---either by $Z^0$-boson exchange between the electrons and the nucleus or by {\P}-violating inter-nuclear forces.
It is manifested in {\P}-violating atomic observables, the measurement of which provide a unique and effective channel for probing the Standard Model (SM) and searching for physics  beyond it.

Experiments in Cs have been the focus of much of the attention over the past few decades. 
The exceptionally precise measurement of the Cs $6S$--$7S$ PNC amplitude~\cite{Wieman1997}, in conjunction with the highly-accurate calculations required for the interpretation (see \cite{DzubaCPM1989plaPNC,Blundell1990,KozlovCs2001,DzubaCs2002,\PBD,OurCsPNC2012} and references therein), led to a determination of the \el{133}Cs nuclear weak charge ({\Q}), an electron--nucleus weak coupling constant, that stands as the most precise low-energy test of the SM to date. 
The result of this analysis differers from the SM prediction by 1.5$\sigma$~\cite{OurCsPNC2012}.
Though this should be considered reasonable agreement, it does indicate that further investigations may yield important new results.

Much of the interest in the area of atomic PNC has been focussed on several other important areas: measuring PNC in a chain of isotopes~\cite{DzubaEnhance1986};  nuclear anapole moments (AMs)~\cite{Flambaum1980,Flambaum1984} (see also \cite{Haxton2001});
and PNC in molecules~\cite{Sushkov1978}.
Accurate atomic calculations are not required for interpreting the measurements of PNC in a chain of isotopes of the same atom, since the atomic structure remains largely unchanged and cancels in the ratio.
The nuclear AM, first introduced by Zel'dovich~\cite{Zeldovich1958}, is a {\P}-violating, {\T}-conserving nuclear moment borne of {\P}-violating forces inside the nucleus.
The experiment~\cite{Wieman1997} of the Wieman group provides the only observation of a nuclear AM so far; further measurements of AMs would provide especially valuable information for the study of hadronic parity violation.

It should be noted that any ``new physics'' involved in atomic PNC would constitute a relatively small correction to an already very small effect.
The case of electric dipole moments (EDMs), however, is somewhat different.
Permanent EDMs of fundamental particles---which are necessarily {\P}- and {\T}-violating---are highly suppressed in the SM, and  those predicted by new theories are often many orders of magnitude larger.
Atomic and molecular EDMs are therefore particularly sensitive probes for theories beyond the SM.
Recent advances in both theoretical and experimental techniques makes this a very exciting area for potential discovery in the near future, e.g., in constraining the electron EDM.
Furthermore, if {\C\P\T} is a good symmetry (as it is in gauge theories), {\T}-violation would be accompanied by {\C\P}-violation, which was first observed in 1964 in the decay of $K^0$ particles \cite{Christenson1964}.
It is well known that the {\C\P}-violation allowed by the SM is insufficient to explain the matter-antimatter asymmetry of the universe; the search for new sources of {\T}-  and {\C\P}-violation is therefore a crucial front for fundamental physics.

We also discuss recent proposals to search for the parity and time-reversal violating effects that are induced in atoms and molecules via their interaction with dark matter, including axions.
The considered effects are linear in the small parameter that quantifies the interaction strength between dark matter and ordinary matter particles; most current techniques search for effects that are at least quadratic in this parameter.

\section{MANIFESTATION OF ATOMIC PARITY VIOLATION}\label{sec:pnc}

\subsection{Sources of Atomic Parity Violation}

The Hamiltonian describing the electron-nucleus weak interaction due to $Z^0$-boson exchange can be expressed
\begin{equation}
\hat h_{\rm PNC} = \frac{-G_F}{\sqrt{2}}\sum_N\left(
	C_{1N}\bar{N}\g_\mu N  \bar e \g^\mu \g_5 e
+	C_{2N}\bar{N}\g_\mu \g_5 N  \bar e \g^\mu e
\right),
\label{eq:int-pnc}
\end{equation}
where the sum runs over all nucleons, $e$ and $N$ are the electron and nucleon {\wf}s, respectively, $G_F\simeq1.166\E{-5} \si{ GeV}^{-2}$ is the Fermi weak constant,
$\gamma_\mu$ and $\g_5$ are Dirac matrices,
 and to lowest order in the SM,
$C_{1n} = -1/2$,
$C_{1p} = (1-4\sin^2\theta_W)/2\approx0.04$ (where $n$ and $p$ denote neutrons and protons),
and
$C_{2p}=-C_{2n}=(1-4\sin^2\theta_W)\lambda/2 \approx0.05$, where $\lambda\approx1.26$, and $\theta_W$ is the Weinberg angle, $\sin^2\theta_W\approx0.24$.
This Hamiltonian is {\P}-violating, but {\T}-conserving.

Treating the nucleons nonrelativistically, the temporal component ($\mu=0$) of the pseudovector electron (vector nucleon) part of the interaction (\ref{eq:int-pnc}) leads to the nuclear-spin-independent (NSI) Hamiltonian,
\begin{align}
\hat h_{\rm NSI} 
&= \frac{-G_F}{2\sqrt{2}}\Big(
	\Q\tilde\rho(\v{r}) +\left[NC_{1n}- ZC_{1p}\right]  \Delta\rho(\v{r}) 
\Big)\g_5,
\label{eq:h-nsi-th}
\end{align}
where 
$N$ and $Z$ are the number of neutrons and protons, respectively,
$\Q=2ZC_{1p}+2NC_{1n}\approx-N$, and
$\tilde\rho=(\rho_n+\rho_p)/2$ and 
$\Delta\rho=(\rho_n-\rho_p)$
with
$\rho_{n,p}$ the normalized nucleon density.
In the calculations, it is assumed that $\rho_n=\rho_p=\rho$, and the second term in (\ref{eq:h-nsi-th}) drops out.
In reality, there is a small difference between average radii of protons and neutrons, the so-called neutron skin.
Though small, this gives an important correction that will be discussed in the coming sections.

The spatial components of the vector electron part of 
(\ref{eq:int-pnc}) lead to the nuclear-spin-dependent (NSD) Hamiltonian 
\begin{align}
\hat h_{\rm NSD}^{Z} 
&= \frac{-G_F}{\sqrt{2}}  \kappa_Z  \frac{K-1/2}{I(I+1)}  \v{\a}\cdot\v{I}   \rho(\v{r}),
\label{eq:h-nsd-Z}
\end{align}
where
$\v{\a}=\g_0\v{\g}$,
 $\kappa_Z=-C_{2n,p}$, 
and
$K=(I+1/2)(-1)^{I+1/2-l}$ with $l$ the orbital momentum of the unpaired nucleon.
This contribution is suppressed due to a number of factors;
the coefficient $|C_{2N}|\ll |\Q|$, and also (unlike in the NSI case) the nucleons do not contribute coherently. In the shell model only the valence (unpaired) nucleons contribute. 
There is also a NSD contribution that comes from the interaction with {\Q} perturbed by the hyperfine interaction \cite{Flambaum1985a,Bouchiat1991},
\begin{equation}
\hat h_{\rm NSD}^{Q} = \frac{G_F}{\sqrt{2}}  \kappa_Q  \frac{\v{\a}\cdot\v{I}}{I}  \rho(\v{r}),
\label{eq:h-nsd-Q}
\end{equation}
which is suppressed by the ratio of hyperfine to fine-structure coefficients:
$\kappa_Q=-\frac{1}{3}\Q\frac{\a \mu_N}{m_p R_N}\simeq2.5\E{-4}A^{2/3}\mu_N$ ($A=N+Z$, $m_p$ is the nucleon mass, $\alpha$ is the fine-structure constant, $R_N$ is the nuclear radius, and $\mu_N$ is the nuclear magnetic moment).

For heavy atoms, however, it is the contribution from the AM  of the nucleus that dominates the NSD effects.
The Hamiltonian describing the interaction of atomic electrons with the nuclear AM is
\begin{equation}
\hat h_{\rm NSD}^{a} = \frac{G_F}{\sqrt{2}}  \kappa_a  \frac{K}{I(I+1)}  \v{\a}\cdot\v{I}  \rho(\v{r}),
\label{eq:hanm}
\end{equation}
where $\kappa_a\sim\a A^{2/3}$ for heavy atoms.
The investigation of AMs will be discussed further in Sec.~\ref{sec:anm}.

\begin{figure}[t!]
\centering
\includegraphics[width=0.327\textwidth]{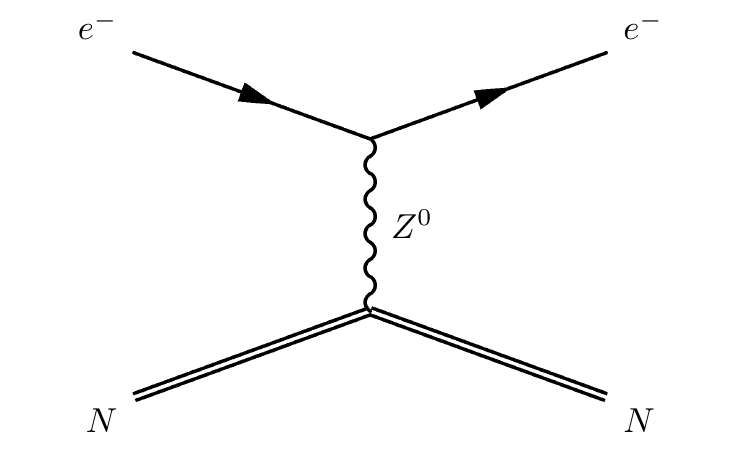}
\includegraphics[width=0.327\textwidth]{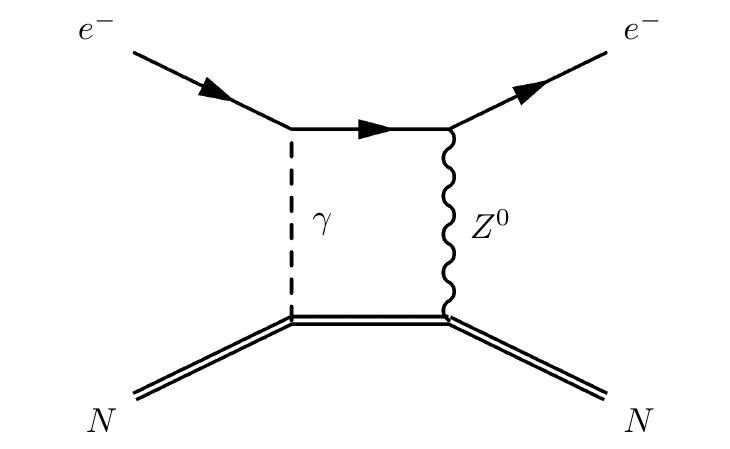}
\includegraphics[width=0.327\textwidth]{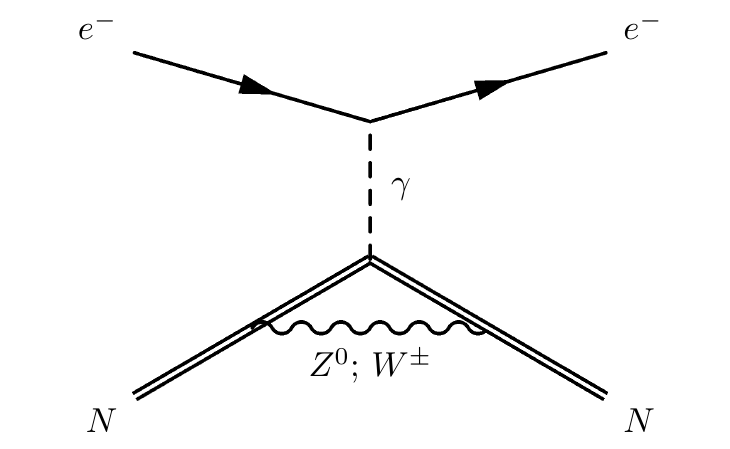}
\caption{ \small
	Example diagrams representing the interaction with
	{\Q} (\ref{eq:h-nsi-th}) and (\ref{eq:h-nsd-Z}),
	{\Q} perturbed by the hyperfine interaction (\ref{eq:h-nsd-Q}), 
	and the nuclear AM (\ref{eq:hanm}), respectively.
}
\label{img:pnc-feyn}
\end{figure}


Example diagrams for the contributions to atomic PNC are represented in Fig.~\ref{img:pnc-feyn}.
Overall, the PNC Hamiltonian can be written as the sum the the NSI and NSD parts,
\begin{equation}
\hat h_{\rm PNC} = 
\hat h_{\rm NSI} + \hat h_{\rm NSD}=
\frac{G_F}{\sqrt{2}} \left(
	\frac{-\Q}{2}\g_5+
	\varkappa\frac{\v{\a}\cdot\v{I}}{I}
 \right)\rho(\v{r}),
\label{eq:hpnc}
\end{equation}
where
$\varkappa=\frac{K}{I+1}\kappa_a-\frac{K-1/2}{I+1}\kappa_Z+\kappa_Q$.
The contributions from the NSI and NSD parts have different experimental signatures, and can thus be treated separately in the analysis. 
It should also be noted that the NSI part is a scalar interaction, and therefore cannot mix atomic states of different angular momentum $J$, whereas the vector NSD interaction can 
($\Delta J\leq 1$, $J_i+J_f>0$).

\subsection{Measurements and Calculations of Atomic PNC}

The prospect of measuring PNC in atoms was first considered for H in 1959 by Zel'dovich~\cite{Zeldovich1959}, who concluded that the effect was too small to be measurable.
More than a decade later, however, the Bouchiats demonstrated that the magnitude of atomic PNC scales a little faster than $Z^3$~\cite{Bouchiat1974a,Bouchiat1974,Bouchiat1975}, where $Z$ is the nuclear charge, meaning that there was a real possibility for non-zero measurements in heavier systems.
See also the book \cite{Khriplovich1991} and the earlier review \cite{GingesRev2004}.

From quantum electrodynamics (QED), an electric dipole ($E1$) transition between atomic states of the same parity cannot arise without external fields due to the conservation of parity.
However, the weak interaction, which violates parity, leads to the mixing of opposite-parity states and therefore gives rise to small {\P}-violating $E1$ amplitudes between states ($a\to b$) of the same (nominal) parity, known as PNC amplitudes:  
\begin{equation}
E_{\rm PNC}^{a\to b} = \sum_n \left[
	\frac{\bra{b}\v{d}\ket{n}\bra{n}\hat h_{\rm PNC}\ket{a}}{E_a-E_n} +
	\frac{\bra{b}\hat h_{\rm PNC}\ket{n}\bra{n}\v{d}\ket{a}}{E_b-E_n}
 \right],
\label{eq:epnc}
\end{equation}
where $\v{d}$ 
is the operator of the $E1$ interaction.
In experiments, it is typically the interference of this amplitude with a {\P}-conserving effect that is directly measured.
In the case of Stark-interference experiments, such as that used for Cs~\cite{Wieman1997}, the {\P}-conserving effect is induced by an applied static electric field.
The electric field gives rise to the ``Stark-induced'' $E1$ amplitude, $E_{\rm Stark}$, which is proportional to the electric field strength, $\mathcal{E}$, and the vector transition polarizability, $\beta$: 
$E_{\rm Stark}\sim \mathcal{E}\beta$. 
The ratio ${\rm Im}(E_{\rm PNC})/\beta$ is measured; as such, in order to extract the amplitude $E_{\rm PNC}$, a determination of $\beta$ is also required.

Stark interference is not the only method that has been successfully utilized.
The first observation of atomic PNC was made in 1978 at Novosibirsk using the ``optical rotation'' technique with Bismuth \cite{Barkov1978}.
Such experiments aim to measure the interference between the {\P}-violating $E_{\rm PNC}$ and {\P}-conserving $M1$ transitions between the same states.
This relies on the fact that PNC in atoms produces a ``spin helix'' (see, e.g., \cite{Khriplovich1991}), which interacts differently with left- and right-polarized light. 
The plane of polarization of light is rotated as the light passes through an atomic vapour b\cite{Zeldovich1959}.
The angle of rotation for light that is tuned to a highly forbidden transition ($a\to b$),  is proportional to ${\rm Im}(E_{\rm PNC})/\bra{b}M1\ket{a}$, and it is this quantity that measured.

\begin{table}[t!]
\centering
\caption{Summary of the more recent/accurate measurements of atomic PNC.}
\begin{tabular}{ll D{,}{}{2.6}  lll}
\toprule
\toprule
\multicolumn{2}{c}{System} & 
\multicolumn{1}{c}{$-\text{Im}(E_{\rm PNC})/M1_{ab}$ ($10^{-8}$)} & 
\multicolumn{1}{c}{Year} & 
\multicolumn{2}{c}{Source} \\ 
\midrule
\el{209}Bi	&	${}^4S_{3/2}$--${}^2D_{3/2}$	
		&	10,.12(20)	&	1991	&	Oxford	&	\cite{Stacey1991}\\
			&	${}^4S_{3/2}$--${}^2D_{5/2}$	
		&		9,.8(9)	&	1993	&	Oxford	&	\cite{Warrington1993}	\\
\el{208}Pb	&	${}^3P_0$--${}^3P_1$	
		&	9,.86(12)	&	1993	&	Seattle	&	\cite{FortsonPb1993}\\
&
		&	9,.80(33)	&	1996	&	Oxford	&	\cite{Phipp1996}\\
\el{205}Tl	&	$6P_{1/2}$--$6P_{3/2}$	
		&	14,.68(17)	&	1995	&	Seattle	&	\cite{Vetter1995}\\
&
		&	15,.68(45)	&	1995	&	Oxford	&	\cite{Edwards1995}\\
\bottomrule
\toprule
\multicolumn{2}{c}{System} & 
\multicolumn{1}{c}{$-\text{Im}(E_{\rm PNC})/\beta$ (mV/cm)} & 
\multicolumn{1}{c}{Year} & 
\multicolumn{2}{c}{Source} \\ 
\midrule
\el{174}Yb	&	${}^1S_{0}$--${}^3D_{1}$	
		&	39,(6)	&	2009	&	Berkeley	&	\cite{Tsigutkin2009}\\
\el{133}Cs	&	$6S_{1/2}$--$7S_{1/2}$	
		&	1,.5935(56)	&1997	&	Boulder	&	\cite{Wieman1997}	\\
&
		&	1,.538(40)	&2005	&	Paris		&	\cite{Guena2003,Guena2005}	\\
\bottomrule
\bottomrule
\end{tabular}
\label{tab:pnc-meas}
\end{table}%

Since then, PNC has also been successfully observed in Pb, Tl, Yb, and Cs.
Table~\ref{tab:pnc-meas} presents a brief summary of some of the more accurate non-zero measurements of atomic PNC and Table~\ref{tab:pnc-calc} presents the corresponding most accurate calculations.
The calculations are presented in units of $10^{-11}i(\Q/N)$, where $N$ is the number of neutrons; this factor is chosen since $\Q\approx -N$.
Theoretical and experimental work has been carried out for many other systems, see Sec.~\ref{sec:newpnc}.

The $Z^3$ scaling of atomic PNC means that heavier atoms are favored for the measurements, since it is natural to expect a higher experimental sensitivity with a larger effect.
However, in order to extract the relevant electroweak parameters, highly accurate atomic calculations are required, which work best for simpler atoms.
Cesium, with high nuclear charge, $Z=55$, and relatively simple electron structure (single $6s$ valence electron above tight Xe-like $5p^6$ core) is an ideal compromise between a large effect and simplicity in the calculations.
The incredible precission that has been attained in both the theoretical and experimental determinations of PNC in Cs have made it the focus of much of the research in this area, and has made PNC in Cs one of the most sensitive tests for new physics beyond the SM.

\begin{table}[t!]
\centering
\caption{Most accurate calculations of $E_{\rm PNC}$ {{[{$-10^{-11}i(-\Q/N)$ a.u.}]}} for transitions listed in Table~\ref{tab:pnc-meas}.}
\begin{tabular}{ll D{,}{}{5.9} llll}
\toprule
\toprule
\multicolumn{2}{c}{System} & 
\multicolumn{1}{c}{$E_{\rm PNC}$} & 
\multicolumn{1}{c}{Year} & 
\multicolumn{2}{c}{Source} \\ 
\midrule
\el{209}Bi	&	${}^4S_{3/2}$--${}^2D_{3/2}$	
		&	26,(3)&	1989	& Dzuba {\sl et al.}&\cite{DzubaCPM1989plaPNC} \\
			&	${}^4S_{3/2}$--${}^2D_{5/2}$	
		&	4,(3)	&	1989	&  Dzuba {\sl et al.}&\cite{DzubaCPM1989plaPNC}	\\
\el{208}Pb	&	${}^3P_0$--${}^3P_1$	
		&	28,(2)	&	1988	&	 Dzuba {\sl et al.}&\cite{DzubaBiPb1988}\\
\el{205}Tl	&	$6P_{1/2}$--$6P_{3/2}$	
		&	27,.0(8)	&	1987 &	 Dzuba {\sl et al.}&\cite{Dzuba1987jpb}\\
&
		&	27,.2(7)	&	2001 &	 Kozlov {\sl et al.}&\cite{Kozlov2001}\\
\el{174}Yb	&	${}^1S_{0}$--${}^3D_{1}$	
		&	195,(25)	&	2011	&	Dzuba {\sl et al.}&\cite{Dzuba2011Yb}\\
\el{133}Cs	&	$6S_{1/2}$--$7S_{1/2}$	
		&	0,.8977(40)	& 2012	& Dzuba {\sl et al.}&\cite{DzubaCs2002,OurCsPNC2012}	\\
\bottomrule
\bottomrule
\end{tabular}
\label{tab:pnc-calc}
\end{table}

\subsection{Parity Nonconservation in Cesium}\label{sec:cs}

The possibility of measuring PNC in Cs was first considered in 1974 by the Bouchiats \cite{Bouchiat1974a}, who also made the first observation in 1982~\cite{Bouchiat1982}.
Since then, several independent measurements have been performed by the Paris and Boulder groups, led by M.-A.~Bouchiat and C.~Wieman, respectively.
A summary of the main results is presented in Table~\ref{tab:cs-meas}.

\begin{table}[b!]
\centering
\caption{Measurements of the $6S$--$7S$ NSI-PNC amplitude in \el{133}Cs (mV/cm).}
\begin{tabular}{llll}
\toprule
\toprule
\multicolumn{1}{c}{$-\text{Im}(E_{\rm PNC})/\beta$} & 
\multicolumn{1}{c}{Year} & 
\multicolumn{2}{c}{Source} \\ 
\midrule
1.52(18)	&1982--6		&Paris		&\cite{Bouchiat1986,Bouchiat1982}	\\
1.576(34)	&1985--8	&Boulder	&\cite{Noecker1988,Gilbert1986,Watts1985}	\\
1.5935(56)	&1997		&Boulder	&\cite{Wieman1997}	\\
1.538(40)	&2003--5	&Paris		&\cite{Guena2005,Guena2003}	\\
\bottomrule
\bottomrule
\end{tabular}
\label{tab:cs-meas}
\end{table}

\begin{figure}[t!] 
\centering
\includegraphics[width=0.5\textwidth]{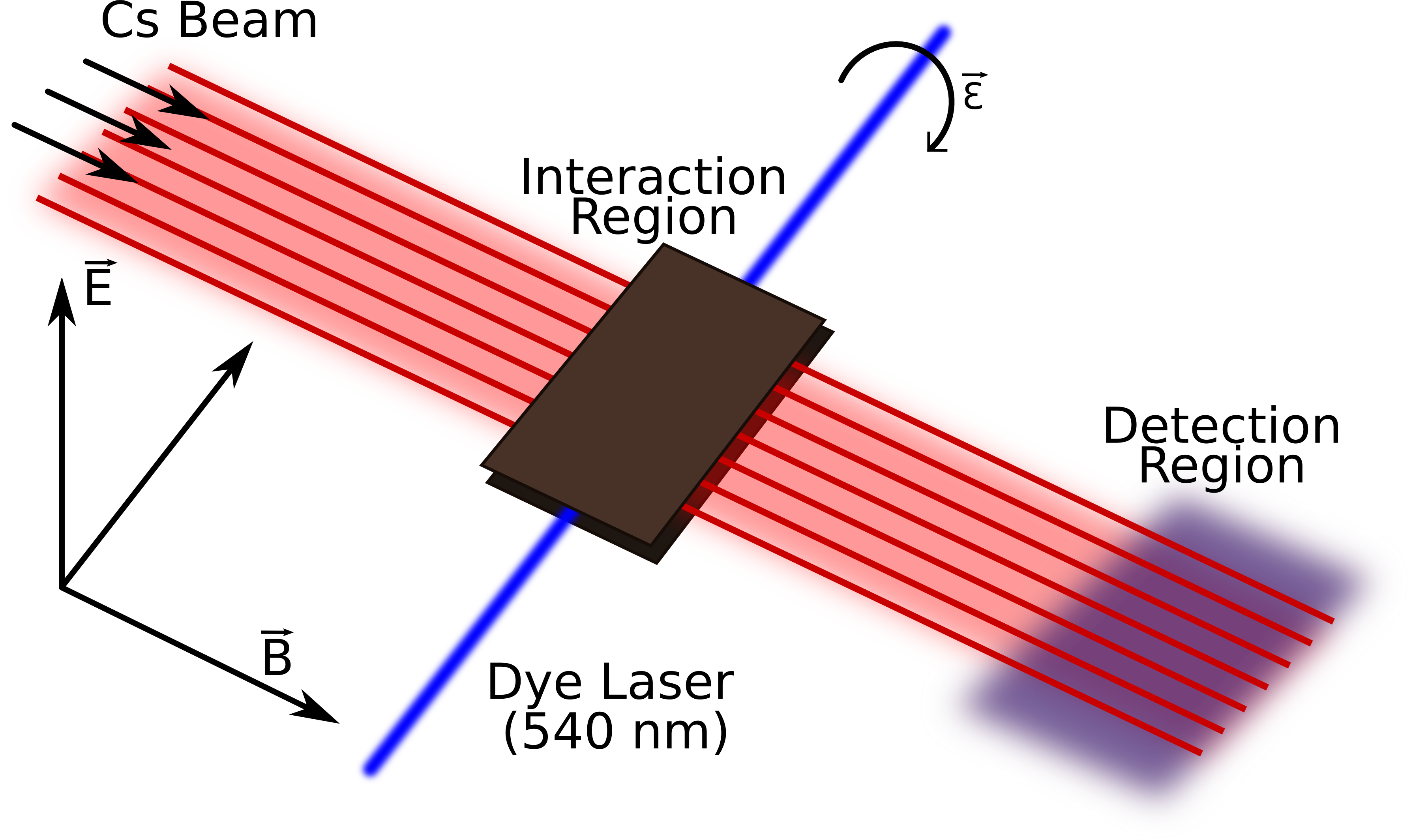}
~~~~~ 
\includegraphics[width=0.3\textwidth]
{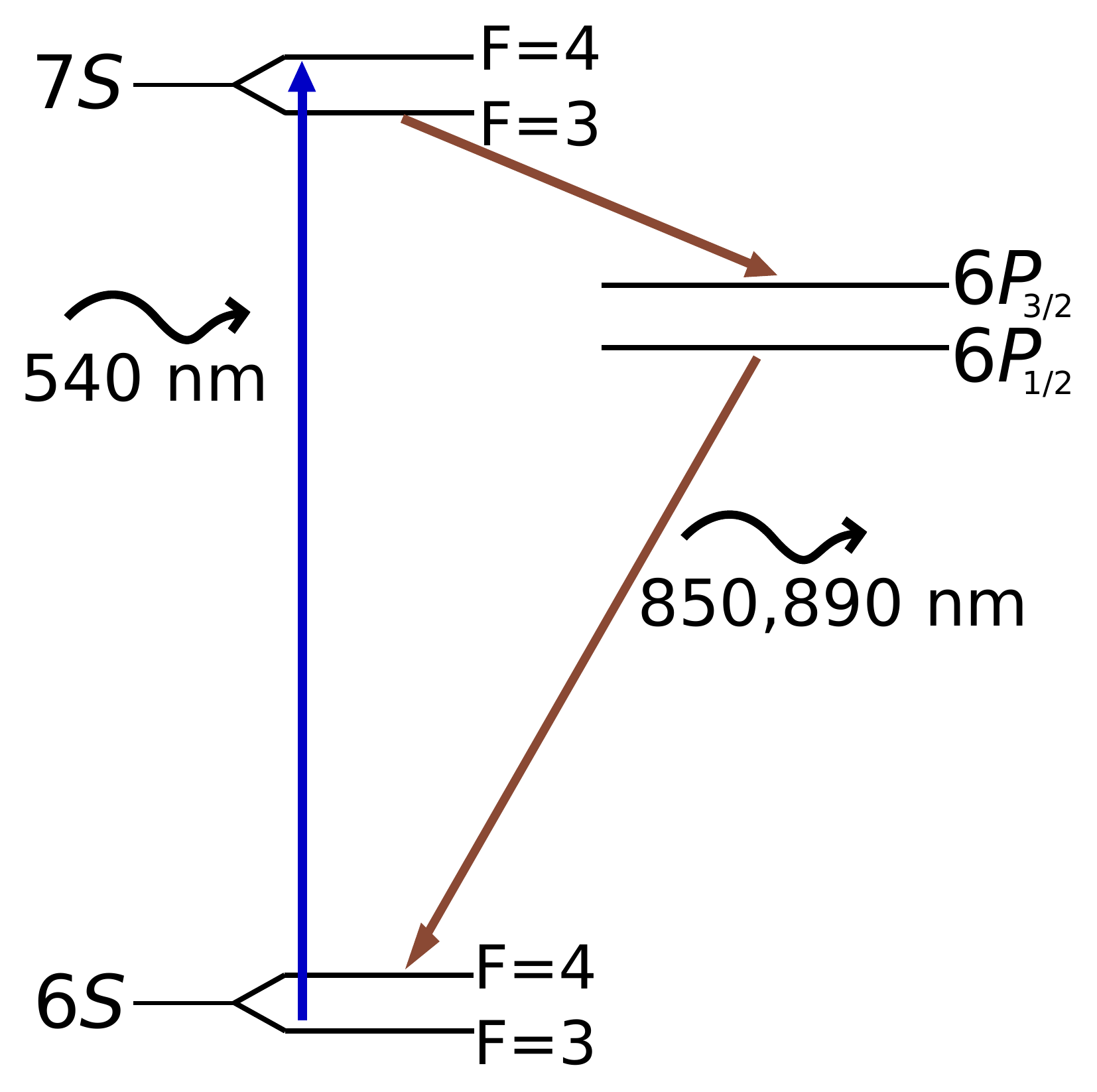}
\caption{\small
	As the beam of Cs passes through the region of perpendicular magnetic and electric fields, the $6S$--$7S$ transition is excited by the dye laser. 
	The transition rate is determined from the intensity of 850 and 890 nm $6P_{1/2,3/2}\to6S$ fluorescence.	
}
\label{fig:cs-app}
\end{figure}

The measurements culminated in 1997 when the Boulder group performed an extraordinarily precise measurement with an uncertainty of just 0.35\% \cite{Wieman1997}, a relative precision unmatched by any other atomic PNC measurement to date. 
They used a Stark-interference technique in which a beam of atomic Cs passes through a region of perpendicular electric, magnetic, and laser fields, as shown in Fig.~\ref{fig:cs-app}.
This process excites the highly forbidden $6S$--$7S$ transition, which contains a small part that is due to the mixing of opposite-parity states by the electron--nucleus weak interaction (\ref{eq:hpnc}).
The transition rate is obtained by measuring the amount of 850- and 890-nm light emitted in the 
$6P_{1/2,3/2}\to6S$ step of the $7S\to6S$ decay sequence.
The {\P}-violating part of the amplitude manifests itself in small modulations to the transition rate as the ``handedness'' of the experiment is changed by reversing the direction of all fields;
see \cite{Wieman1997} and references therein for details.
Their final result was
\begin{equation}
-{\rm Im}\frac{E_{\rm PNC}}{\beta}=
	\begin{cases}
		1.6349(80)   \,{\rm mV/cm}  \quad (6S_{F=4}\to7S_{F=3}) \\
		1.5576(77)   \,{\rm mV/cm}  \quad (6S_{F=3}\to7S_{F=4})  \\
	\end{cases}.
\label{eq:cspnc}
\end{equation}
Since the interaction with the weak charge is independent of nuclear spin, it contributes the same amplitude to each hyperfine component. Thus, by averaging over the hyperfine components, one can determine the contribution due to $\Q$: 
$-{\rm Im}({E_{\rm PNC}}/{\beta})=1.5935(56)\,{\rm mV/cm}$. 
The precision of the Boulder measurement for the first (and so-far only) time also allowed for the detection of NSD-PNC effects, which led to a determination of the \el{133}Cs nuclear AM \cite{Wieman1997,FlambaumAnM1997}.

The more recent measurements of the Paris group \cite{Guena2003,Guena2005} (see also \cite{Lintz2007}) used a different method---chiral optical gain---to detect the PNC signal.
The results using this method are not at the same level of accuracy as the Boulder measurements \cite{Wieman1997}, however, promising progress has been made.
These new results may prove particularly significant as an independent verification of the important Boulder results.
Even more recently, work on developing new methods has been under way; in 2014 an experimental group from Indiana successfully utilized a ``two-pathway coherent control interference'' technique 
to measure an $M1$ transition amplitude in Cs~\cite{Antypas2013b}, which may be applied to measuring PNC in Cs with reduced errors from systematics and unwanted interference~\cite{Antypas2014}.

In order to determine a value of $\Q$ for $^{133}$Cs from the measurement~\cite{Wieman1997}, both a value for the vector transition polarizability, $\beta$, and a calculation of the atomic structure~(\ref{eq:epnc}) are required. 
The most accurate value, 
$  \beta = 26.957(51)\, a_B^3$
($a_B$ the Bohr radius), comes from an analysis~\cite{DzubaCs2000} of the Bennett and Wieman measurements~\cite{Wieman1999}.
This is not the only determination of $\beta$, and less than perfect agreement exists between methods, see, e.g.,~\cite{Cho1997,DzubaCs2000,DzubaCs2002}.
 At the moment, however, this is not a major problem since the uncertainty in the extraction of $\Q$ is dominated by the calculations.
Combined with the most accurate calculations available at the time 
 \cite{DzubaCPM1989plaPNC,Blundell1990}, the measurements indicated good agreement with the SM.
However, the declared theoretical uncertainty of these early calculations (1\%) wasn't at the same level as the measurements.

An analysis of the accuracy of the calculations was performed in light of new experimental tests concerning $E1$ amplitudes and hyperfine constants
in \cite{Wieman1999}. The authors noted that many of the previous discrepancies between theory and experiment were resolved in favor of theory, which led them to conclude that the accuracy of the calculations for Cs~\cite{DzubaCPM1989plaPNC,Blundell1990} was actually as good as 0.4\%~\cite{Wieman1999}.
The new analysis indicated that the observed value for the weak charge of the \el{133}Cs nucleus differed from the SM prediction by 2.5$\sigma$---signalling the possibility that new physics had been observed. 
The excitement was short-lived, however, when the inclusion of the Breit (magnetic and retardation)~\cite{Derevianko2000} and radiative QED corrections (see \cite{\QED,FlambaumQED2005,Shabaev2005a} and references therein) into the calculations led to a triumphant restoration of the Cs results with the SM. 
Concurrently, several new calculations~\cite{KozlovCs2001,DzubaCs2002} agreed well with the previous results~\cite{DzubaCPM1989plaPNC,Blundell1990} and confirmed the suggestion made in \cite{Wieman1999} that the theoretical accuracy was high.
At this point, all recent calculations were in excellent agreement, and the new value of {\Q} was consistent with the SM, being about $1\sigma$ smaller than predicted.%

\begin{table}[b!]
\centering
\caption{Calculations of the \el{133}Cs $6S$--$7S$ PNC amplitude {{[{${10^{-11}i(\Q/N)}$ a.u.}]}}. Breit, QED, and neutron skin corrections are not included.  
}
\begin{tabular}{D{.}{.}{1.5}lll|D{.}{.}{1.7}lll}
\toprule
\toprule
\multicolumn{1}{c}{$E_{\rm PNC}$} & 
\multicolumn{1}{c}{Year} & 
\multicolumn{2}{c}{Source} &
\multicolumn{1}{|c}{$E_{\rm PNC}$} & 
\multicolumn{1}{c}{Year} & 
\multicolumn{2}{c}{Source} \\ 
\midrule
0.88(3)		&1984		& Dzuba {\sl et al.}	&\cite{DzubaPNC1984}	&
0.905(9)	&2001		& Kozlov {\sl et al.}	&\cite{KozlovCs2001}			\\
0.90(2)		&1987		& Dzuba {\sl et al.}	&\cite{Dzuba1987ps}	&
0.9078(45)	&2002		& Dzuba {\sl et al.}	&\cite{DzubaCs2002}			\\
0.95(5)		&1988		& Blundell {\sl et al.}&\cite{JohnsonPNC1988}			&
0.8998(24)	&2009		& Porsev {\sl et al.} 	&\cite{\PBD}					\\
0.908(9)	&1989		& Dzuba {\sl et al.}	&\cite{DzubaCPM1989plaPNC}	&
0.9079(40) &2012		&Dzuba {\sl et al.}	&\cite{OurCsPNC2012}		\\ 
0.909(9)	&1990		& Blundell {\sl et al.}&\cite{Blundell1990}		& &&&\\
\bottomrule
\bottomrule
\end{tabular}\\
\label{tab:cs-comp}
\end{table}

More recently, however, the situation changed when a new calculation was reported by Porsev {\sl et al.}~\cite{\PBD}. 
They used a very sophisticated approach, applying the coupled-cluster method with single, double, and valence triple excitations (CCSDvT)---for details, see \cite{\PBD} and references therein.
Claiming just 0.27\% uncertainty of the calculations, their ``correlated'' PNC amplitude (not including Breit, QED or neutron-skin corrections) was about 0.9\% smaller than the results of previous calculations; see Table~\ref{tab:cs-comp}. 
This led to perfect agreement with the SM; the central points for the weak nuclear charge extracted from the measurements coincided exactly with that predicted by the SM:
$\Q = -73.16(29)_{\rm exp}(20)_{\rm th}$, 
$\Q^{\rm SM} = -73.16(3)$\footnote{Note that the SM prediction has since been updated: $\Q^{\rm SM}(^{133}{\rm Cs}) = -73.23(2)$~\cite{PDG2012}.}~\cite{\PBD}.
The variation from the previous calculations was attributed to the role of higher-order correlations.

In \cite{\PBD}, the ``main'' ($n=6,7,8,9$) terms in the summation (\ref{eq:epnc}) were treated very accurately with the CCSDvT method; however, a significantly less accurate method was used to calculate the remaining core ($n \leq5$) and highly excited ``tail'' ($n>9$) terms.
The main terms contribute about 97\% to the total amplitude, though at this level of precision accuracy of the remaining terms is important also.
From an analysis of the variation of these terms in different approximations, a 10\% uncertainty for the core and tail was adopted.
In \cite{OurCsPNC2012}, however, it was shown the inclusion of many-body effects (correlations and core polarization) that were neglected in \cite{\PBD} for these terms has a significant impact on the calculations. 
With a change in sign, the core contribution shifts by about 200\%; far beyond the declared 10\% uncertainty.
The tail contribution also becomes significantly larger.
With the core and tail contributions of \cite{\PBD} substituted by those calculated in \cite{OurCsPNC2012}, the excellent agreement with previous calculations is restored.

The final result
 from \cite{OurCsPNC2012} 
(last row in Table \ref{tab:cs-comp}) 
leads to a value of
$\Q = -72.58(29)_{\rm exp}(32)_{\rm th}$, 
which is in reasonable agreement with the SM prediction. 
Adding theoretical and experimental errors in quadrature, the Cs PNC result deviates from
the SM value by 1.5$\sigma$:  
$\Delta \Q  \equiv \Q-\Q^{\rm SM} = 0.65(43).$  
This can be related to the deviation in $\sin^2\theta_W$, giving 
$\sin^2\theta_W = 0.2356(20)$, $1.5\sigma$ from the SM value of $0.2386(1)$~\cite{PDG2012} at near zero momentum transfer.

Though the results of \cite{\PBD} and \cite{OurCsPNC2012} both indicate reasonable agreement with the SM, the constraints on new physics beyond it are significantly different.
New physics originating from vacuum polarization can be described by the weak isospin-conserving $S$ and -breaking $T$ parameters:  
$\Delta \Q= -0.800\,S-0.007\,T$~\cite{Peskin1992,Rosner2002}.
At the 1$\sigma$ level, the result of \cite{OurCsPNC2012} leads to $S=-0.81(54)$, whereas in \cite{\PBD} it was constrained at $|S|<0.45$.
Additionally, a positive $\Delta \Q$ can be interpreted as evidence for an extra neutral boson, $Z{_{\chi}}$, in the weak interaction~\cite{Marciano1990}.
The result of \cite{OurCsPNC2012} leads to a constraint on its mass of
$M_{Z_{\chi}}>650$ GeV/$c^2$ (85\% confidence level),     
a significantly less stringent bound than the $1.4$ TeV/$c^2$ set in \cite{\PBD}.

Furthermore, recent measurements made by the $Q_{\rm weak}$ Collaboration in 2013 at the Jefferson Lab have led to the first determination of the weak charge of the proton, $\Q^p=0.064(12)$~\cite{ProtonQw2013}. 
Combining this with the weak charge obtained via Cs PNC leads to a value for the weak charge of the neutron, $\Q^n=-0.975(10)$.

\section{FUTURE PROSPECTS FOR ATOMIC PNC}\label{sec:future}

\subsection{New Measurements of PNC}\label{sec:newpnc}  

Though it remains the case that the Cs results are the most precise atomic PNC measurements, there are promising signs for successful parity violation determinations in several other atomic systems.
Heavy analogues of Cs, such as Fr, have the advantage that the PNC effects can be largely enhanced  \cite{DzubaFr1995,Safronova2000}.  
Preparations for PNC experiments in Fr are currently under way at the TRIUMF facility in Vancouver~\cite{\triumf}.

The largest PNC signal to date was observed in Yb at Berkeley~\cite{Tsigutkin2009}.
The effect was about two orders of magnitude larger than that of Cs, and significant improvements in the sensitivity are expected in the near future \cite{Tsigutkin2010,Dounas-Frazer2011}.
Though the accuracy of the interpretation for Yb is only around 10\% \cite{Porsev2000,DeMille1995,Dzuba2011Yb}, it may prove especially fruitful for measurements of the AM and PNC in a chain of isotopes.
Also at Berkeley are ongoing measurements to search for PNC in Dy \cite{Nguyen1997,Leefer2014}.
Dy possesses two nearly degenerate states ($A$ and $B$) of opposite parity and the same angular momentum, $J=10$, at $E = 19797.96$ cm$^{-1}$.
By observing time-resolved quantum beats between these levels caused by interference between the Stark and PNC mixing, the weak-interaction matrix element was found to be
$\bra{A}\hat h_{\rm PNC}\ket{B}=2.3(29)_{\rm stat.}(07)_{\rm sys.}$ a.u.~\cite{Nguyen1997}, consistent with theory \cite{DzubaDy2010}. The unfortunate smallness of the relevant matrix element is due to the fact that the PNC interaction cannot mix the dominant configurations of the $A$ and $B$ states.
Experimental work is continuing, however, with an expected improvement in the statistical sensitivity of a few orders of magnitude~\cite{Leefer2014}; this would provide an important test of the SM and potentially lead to a measurement of the Dy AM.

There have also been suggestions put forward to measure PNC in $S$--$D_{3/2}$ transitions of single-trapped ions, such as Ba$^+$ and Ra$^+$~\cite{Fortson1993} (see also \cite{Pal2009,Wansbeek2008,DzubaPNCsd2001}) and heavier Cs- and Fr-like ions \cite{RobertsActinides2013,Robertssd2014}, which have electron structure similar to Cs. 
Experimental work is currently in progress for Ba\+ at Seattle \cite{Williams2013}, and for Ra\+ at the KVI institute in Groningen~\cite{\kvi}.
By exploiting PNC effects in heavy alkali-like ions, very high experimental sensitivity can be achieved while not impacting too heavily the accuracy of the calculations.
However, the accuracy of the interpretation is unlikely to outperform Cs due to the larger
correlations associated with the $d$-states (only $s$ and $p$ states are involved for Cs) and larger relativistic (QED, Breit) corrections.%

Another possibility is to move towards the lighter elements.
 Rb, a lighter analogue of Cs, is a promising option to search for both {\Q} and the AM~\cite{OurRb2012}. A similar proposal has been put forward for Sr\+ \cite{Dutta2014a}.
  The atomic physics calculations for Rb can surpass the accuracy of those for Cs, due to the simpler electron structure and smaller relativistic corrections.  
This is important, since currently (in Cs) it is the theoretical uncertainty that outweighs the experimental error.

As the theoretical accuracy for the calculations approaches the level already attained in Cs, smaller, sub-1\% corrections become important.
As discussed above, QED effects have already proven to be crucial for the interpretation of the results in Cs. 
Radiative QED corrections to the $\hat h_{\rm NSI}$ matrix elements were calculated in \cite{\QED}.
In \cite{FlambaumQED2005}, the ``radiative potential'' method was developed as a simple yet accurate way of including these effects into the atomic calculations for many-electron systems.
This allows the inclusion of QED effects into the $E1$ matrix elements and energy denominators  [see (\ref{eq:epnc})].
Along with the calculations for the $\hat h_{\rm NSI}$ matrix elements from \cite{\QED}, this method was used to determine QED corrections to several PNC amplitudes in \cite{RobertsQED2013}.
It is crucial that high theoretical accuracy can be confirmed through independent calculations by several different groups. 
At the moment, however, for certain systems there exists small, but significant, disagreement.
A recent study of one particular many-body effect, the so-called {double core polarisation}, that may have been missed in some calculations suggests that this effect has the potential to resolve some of the disagreement in the literature~\cite{RobertsDCP2013}.

Recent developments in experimental techniques are also showing great promise for a highly precise new measurement of atomic PNC in the near future.
For example, in \cite{Bougas2012}, an optical cavity was developed that can enhance optical rotation signals by as much as four orders of magnitude.
This advantage can be made more significant by combining this signal enhancement with further PNC enhancements in diatomic molecules, which have nearby opposite parity states \cite{Sofikitis2014}.

\subsection{PNC in a Chain of Isotopes}\label{sec:coi}

Since the matrix elements of $\hat h_{\rm NSI}$ (\ref{eq:hpnc}) are proportional to the nuclear weak charge, one can express the NSI-PNC amplitude in the form
$E_{\rm PNC} = A \Q$,
where $A$ is an electron structure coefficient.
As discussed above, in order to extract experimental values for {\Q}, highly accurate calculations of $A$ are required, a fact that limits the applicable systems to those with simple electron structure. 
Noting, however, that the electron structure does not depend significantly on the isotope used,
an alternative method was put forward in \cite{DzubaEnhance1986}.
By measuring PNC in the same transition for at least two isotopes of the same atom and taking the ratio, 
\begin{equation}
\mathcal{R} =\frac{E'_{\rm PNC}}{E_{\rm PNC}} = \frac{\Q'}{\Q},
\label{eq:coi}
\end{equation}
the coefficients $A$ cancel, eliminating the need for atomic calculations.
This is referred to as the chain of isotopes method.

The shortcoming of this method, however, is that the neutron distribution [$\rho$ in (\ref{eq:hpnc})] does in fact change slightly between isotopes.
In \cite{Fortson1990} it was noted that possible constraints on new physics derived from measurements of PNC in a chain of isotopes are sensitive to the uncertainties in the neutron distribution, known as the neutron skin, 
which are large enough to be a strong limiting factor for this technique.
To circumvent this problem, it was suggested in \cite{Derevianko2002} that available experimental data on neutron distributions could be used to reduce the uncertainties.
In a more recent work \cite{Brown2009}, the authors performed nuclear calculations and demonstrated that the neutron distributions are actually correlated for different isotopes. 
This means that much of the relevant uncertainty in the ratio (\ref{eq:coi}) is cancelled, and provides a framework for estimating the remaining error.
The conclusion from \cite{Brown2009} is that chain of isotope measurements are in fact markedly sensitive to new physics.
Experiments with the aim of measuring PNC in a chain of isotopes are under way for Dy  \cite{Leefer2014}, Yb \cite{Tsigutkin2009,Tsigutkin2010}, Fr \cite{\triumf},
and 
Ra$^+$ \cite{\kvi}.

\subsection{Nuclear Anapole Moments}\label{sec:anm}

It was first shown by Zel'dovich in 1957 that {\P}-violation inside a charge distribution could give rise to an AM \cite{Zeldovich1958}.
It was subsequently pointed out that an AM in the nucleus would contribute to NSD PNC in atoms and molecules \cite{Flambaum1980} (see also \cite{Flambaum1984,Sushkov1984,Flambaum1985});
in fact it was demonstrated that the effect of the AM was enhanced in heavy nuclei as $A^{2/3}$, and that it dominates the NSD contribution to PNC in atoms and molecules.
This meant that, with sufficiently precise measurements, atomic experiments could be used to study {\P}-violation in the hadron sector.
This ``tabletop'' nuclear physics provides a unique low-energy probe for physics that is relatively inaccessible by other means; see, e.g.,~\cite{Haxton2001,Khriplovich2004}.

\begin{figure}[b!]
\centering
\includegraphics[width=0.3\textwidth]{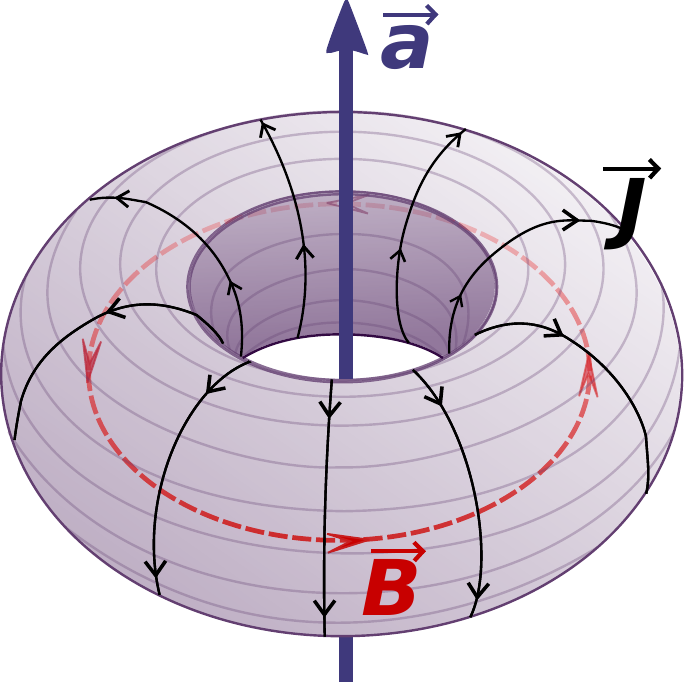} 
\caption{\small
Diagram showing the toroidal current, $\v{\vec{J}}$, the magnetic field it produces, $\v{\vec{B}}$, and the resulting anapole moment, $\v{\vec{a}}$.
}
\label{img:am}
\end{figure}

{\P}-violating forces acting between nucleons create a spin helix structure inside the nucleus.
A part of the vector potential created in this configuration is of a contact nature,
$\v{A}^{\rm AM}=\v{a}\delta^3(\v{r})$, where
$\v{a}=-\pi\int r^2\v{j}(\v{r})\,{\rm d}^3r$
is the AM with $\v{j}(\v{r})$ the electromagnetic current density; see, e.g.,~\cite{Khriplovich1991,Haxton2002}. A diagram of a current distribution that gives rise to an AM is shown in Fig.~\ref{img:am}.
Note that such a moment must violate parity; the AM contains the current vector $\v{j}$, which is {\P}-odd, but it's also directed along the nuclear spin $\v{I}$, which is {\P}-even:  $\braket{\v{a}}=-\pi\braket{r^2\v{j}}=|\v{a}|\v{I}/I$.

The AM is quantified by the dimensionless parameter $\kappa_a$,
\begin{equation}
\v{a}=\frac{1}{e}\frac{G_F}{\sqrt{2}}\frac{K\v{I}}{I(I+1)}\kappa_a,
\end{equation}
where $e$ is the proton charge, and $K$ is defined in (\ref{eq:h-nsd-Z}).
The interaction of atomic electrons with the AM, which has the form 
$\hat h^a=e\v{\a}\cdot\v{a}\rho(\v{r})$ [see (\ref{eq:hanm})], leads to NSD-PNC effects in atoms.
The interaction with the AM is the dominant NSD contribution to PNC in heavy atoms, however its effect is indistinguishable from that of $\kappa_Z$ and $\kappa_Q$ (see Sec.~\ref{sec:pnc}), and these must be calculated and subtracted in order to extract $\kappa_a$ from the measurements.

The Cs experiment~\cite{Wieman1997} of the Boulder group provides the only definitive observation of a nuclear AM to date (there are limits on the \el{203,205}Tl AM from experiment \cite{Vetter1995}). 
The NSD contribution to the $6S$--$7S$ PNC amplitude was found to be 
${\rm Im}(E_{\rm PNC})/\beta=0.077(11)$ (\ref{eq:cspnc}).
In \cite{FlambaumAnM1997}, a value for the NSD-PNC constant
$\varkappa(\el{133}{\rm Cs})=0.393(56)$ 
was extracted from the measurements by taking the ratio of the calculated NSD-PNC amplitude from \cite{Kraftmakher1988} to the NSI amplitude, calculated in \cite{Dzuba1987jpb}. 
These works were chosen since they were performed using an identical technique, and the theoretical uncertainties would cancel in the ratio \cite{FlambaumAnM1997}.
This value was confirmed in \cite{Johnson2003}.

In the single-particle approximation, $\kappa_Z$ for \el{133}Cs is given by $\kappa_Z=-C_{2p}\simeq-0.05$. 
Taking nuclear many-body effects into account, a value of $\kappa_Z=-0.063$ was calculated in \cite{Haxton2001c}.
Atomic calculations performed in \cite{Johnson2003} (see also \cite{Flambaum1985a,Kozlov1988,Bouchiat1991} determined the value $\kappa_Q=0.017$.
Taking these into account leads to a value for the AM constant \cite{GingesRev2004}
\begin{equation}
\kappa_a(\el{133}{\rm Cs})=0.362(62)
\end{equation}
(see \cite{FlambaumAnM1997} for a discussion of finite-nuclear-size effects).

The SM prediction of $\kappa_a$ is highly dependent on nuclear physics calculations.
For \el{133}{\rm Cs}, it ranges from as high as $\kappa_a=0.36$ in the single-particle approximation down to $\kappa_a=0.11$ depending on how the many-body effects are included, see, e.g.,~\cite{Bouchiat1991,Dmitriev1997,Dmitriev2000,Haxton2001c,Haxton2002} (see also \cite{GingesRev2004} for a discussion).
It must, therefore, be concluded that the Cs results are in reasonable agreement with the SM.

There are, however, discrepancies between weak meson-nucleon coupling constants extracted from the Cs AM and those extracted from hadron scattering experiments; see, e.g., \cite{Haxton2013,GingesRev2004}. 
There is also a problem from the measurements in Tl.
The AM of \el{203,205}Tl has been constrained as $\kappa_a=-0.22(30)$ \cite{Vetter1995,Khriplovich1995}, which is inconsistent both with that predicted by nuclear theory (between $\kappa_a=0.10$ and $0.48$, see, e.g.,~\cite{Bouchiat1991,Dmitriev2000,Haxton2001c,Haxton2002}) and with the Cs results.
It is clear that there is much to be gained from further investigation into this field.

The current status of nuclear physics means that even modestly accurate measurements of the AM  can shed light on important physics. 
Therefore, the extreme precision that is required of the atomic calculations for extracting $\Q$ is not necessary.
This frees the possibility of exploiting favorable conditions in more complicated atoms and molecules where the effect is larger.
Also, it would be extremely beneficial to measure AMs for nuclei with an unpaired neutron (Cs and Tl have unpaired protons).  
In \cite{RobertsCL2014}, calculations were performed for AM (due both to unpaired protons and neutrons) and {\Q} induced PNC amplitudes for several heavy rare-earth and actinide atoms in which the effect is enhanced by the presence of pairs of close opposite-parity levels (see also \cite{FlambaumRa1999,DzubaRa2000}).
Experimental work to measure AMs is in progress at Berkeley for Dy  \cite{Leefer2014,Nguyen1997} and Yb \cite{Tsigutkin2009,Tsigutkin2010}, 
at Heraklion for Xe and Hg \cite{Bougas2012},
and at TRIUMF for Fr \cite{\triumf}.

In \cite{Sushkov1978,Flambaum1985} it was noted that the effect of the AM is strongly enhanced in diatomic molecules due to the mixing of close rotational states of opposite parity, including the mixing of $\Lambda$ or $\Omega$ doublets (see also \cite{Labzovsky1978}). 
The PNC effects produced by {\Q} are not enhanced, meaning it is the AM effect that dominates PNC in molecules.
For a review of {\P}- and {\T}-violation in diatomic molecules, we direct the reader to \cite{Kozlov1995}.

The enhancement of the AM effects in molecules is due to the ability of the AM to mix very closely spaced rotational levels of opposite parity.
After averaging over the electron {\wf}, the effective operator acting on the angular variables may contain three vectors:
the direction of molecular axis $\v{N}$, 
the electron angular momentum $\v{J}$,
and the nuclear spin $\v{I}$. 
The scalar products $\v{N}\cdot\v{I}$ and $\v{N}\cdot\v{J}$ are both {\T}- and  {\P}-odd. 
A {\P}-odd, {\T}-even operator must therefore be proportional to the product
$\v{N}\cdot( \v{J} \times  \v{ I})$, which contains nuclear spin $\v{I}$, so the NSI weak charge cannot contribute. 
 The matrix elements of $\v{N}$ between rotational states 
produce $E1$ transitions in polar molecules. 
Therefore, the interaction of the nuclear AM with molecular electrons mixes close  rotational-hyperfine states of opposite parity.  
The intervals between these levels are around five orders of magnitude smaller than those between  opposite-parity states in atoms, meaning the PNC effects can be around five orders of magnitude larger. 
Further enhancement may be achieved by a reduction of the intervals by an external magnetic field~\cite{Flambaum1985}.
Note that very close levels of opposite parity can also be found in heavy atomic systems, such as the actinide and rare-earth metals \cite{DzubaEnhance1986}; however, this is often at the loss of single-particle $s$-$p_{1/2}$ mixing \cite{DzubaDy2010,RobertsCL2014}, which suppresses the overall effect.

Molecules and molecular ions with $\Sigma_{1/2}$ or $\Pi_{1/2}$ electronic ground states are good candidates for the measurements~\cite{Sushkov1978,Flambaum1985}.
Molecular PNC experiments are currently in progress for BaF at Yale~\cite{Cahn2014} and RaF at KVI \cite{Isaev2010}.
Measurements of the AM in molecules also require electron structure calculations for their interpretation.
A number of calculations have been performed for diatomic molecules of experimental interest; see, e.g.,~\cite{Isaev2010,Borschevsky2012a,Nayak2009,
DeMille2008,Dmitriev1992} and references within.

Recent progress in molecular cooling and trapping techniques have made this area particularly exciting for breakthroughs in the very near future; see, e.g., \cite{Carr2009} and references therein. 
Laser cooling of molecules was first demonstrated experimentally with polar SrF molecules in 2010, where temperatures of a few milli-Kelvin were achieved \cite{Shuman2010}.
In mid-2014, magneto-optical trapping of SrF was demonstrated at a temperature of about 2.5 mK \cite{Barry2014}.
Other schemes, such as those employed recently to cool polyatomic CH$_3$F molecules \cite{Zeppenfeld2012}, are also making promising leeway.
Such techniques will prove exceptionally useful not only in searching for PNC, but also increasingly in the search for permanent EDMs of molecules.

\section{ELECTRIC DIPOLE MOMENTS}\label{sec:edm}

\subsection{Manifestations of $T$ Violation in Atoms and Molecules}\label{sec:atmolEDM}  %

\begin{figure}[b!]
\centering
\includegraphics[width=0.8\textwidth]
{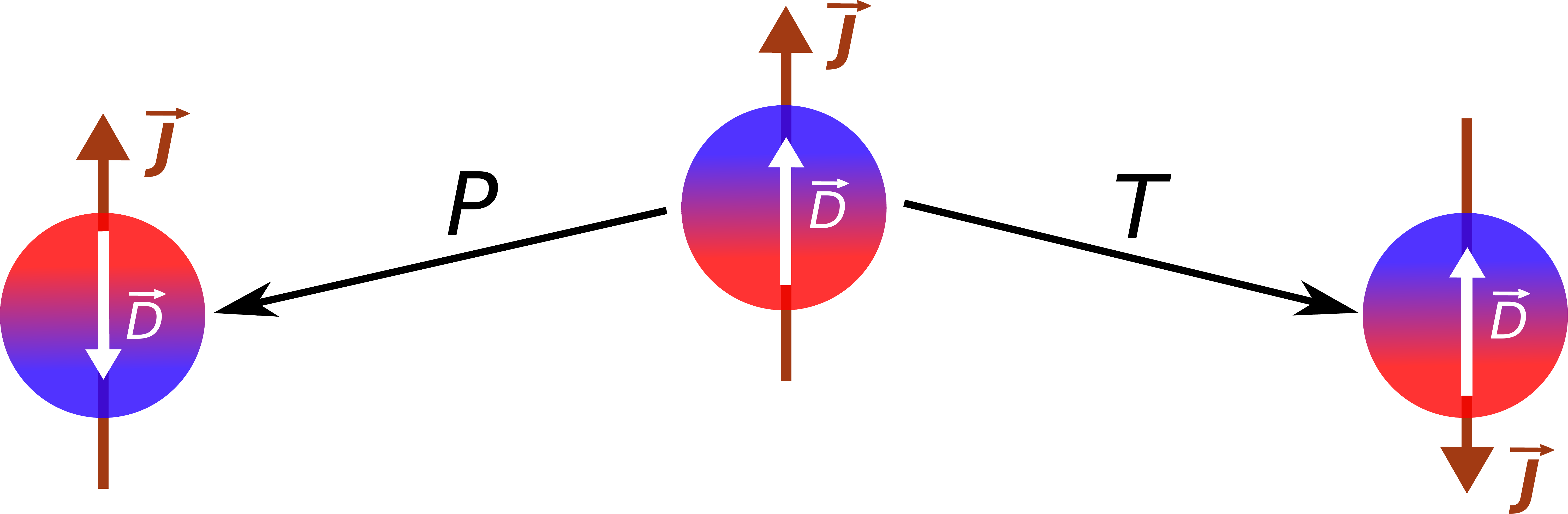}
\caption{\small
The expectation value of the electric dipole operator, \v{d}, lies in the direction of the total angular momentum, \v{J}; however, \v{d} is {\P}-odd and {\T}-even while \v{J} is  {\P}-even and {\T}-odd.
}
\label{img:EDM}
\end{figure}

A permanent EDM of a stable particle (e.g., a nucleon, atom, or molecule) would violate both {\P} and {\T} invariance, see Fig.~\ref{img:EDM}.
\sidebar{Polar molecules}{It is commonly stated that the polar molecules (e.g., H${}_2$O or NH${}_3$) have a permanent EDM; while this is a useful way to describe the molecular interactions, it is not correct. 
In the weak-field limit, the energy shift of such a system is actually quadratic (not linear) in the electric field strength. In fact, this is an example of an induced (i.e.~temporary) EDM, caused by the mixing of near-degenerate opposite-parity states by the electric field, which does not violate {\P} or {\T} invariance; see, e.g.,~\cite{Khriplovich1997}.
}
The SM allows only extremely small EDMs of fundamental particles. 
Conversely, most extensions to the SM predict much larger EDMs, which are within experimental reach---making EDMs an extraordinarily sensitive probe for new physics \cite{Pospelov2005}.
The parameter space for {\C\P}-violation allowed in supersymmetric theories is already very strongly limitted by EDM measurements \cite{Pospelov2005,Engel2013,Chupp2014}.

The EDM, $\v{D}^{(a)}$, of an atom or molecule in state $a$ can arise either from the sum of the intrinsic EDMs of the constituent particles, or from the mixing of opposite-parity states due to a {\P}- and {\T}-violating interaction, $\hat h_{\P\T}$: 
\begin{equation}
\v{D}^{(a)} = 2\sum_n \frac{ \bra{a}\v{d}\ket{n}\bra{n}\hat h_{\P\T}\ket{a}}{E_a-E_n}.
\label{eq:atEDM}
\end{equation}
An atomic or molecular EDM can be generated via several ({\P},{\T})-violating mechanisms, e.g., the interaction with the electron EDM, and ({\P},{\T})-violating electron--nucleon and nucleon--nucleon interactions 
\cite{Sandars1965,Sandars1967}.
Different systems have different sensitivities to the various sources, depending on electronic and nuclear structure. 
For example, in paramagnetic systems (with non-zero $J$), the EDM is due almost entirely to the electron EDM and ({\P},{\T})-violating electron--nucleon interactions.
For diamagnetic systems ($J=0$), however, EDMs are mostly due to the {\T},{\P}-odd inter-nuclear forces and the NSD electron--nucleon interaction.

The existence of ({\P},{\T})-odd nuclear forces gives rise to ({\P},{\T})-violating nuclear moments in the multipole expansion of the nuclear potential.
The lowest-order term in the expansion, the nuclear EDM, is unobservable in neutral atoms due to total screening of the external electric field by the atomic electrons (the Schiff theorem) \cite{Schiff1963}.
We note, however, that it may be possible to observe the nuclear EDM in ions (see, e.g., \cite{KozlovAion2012}).
The first non-vanishing terms that survive the screening in neutral systems are the so-called Schiff moment and the electric octopole moment.
After the nuclear magnetic dipole moment, the lowest magnetic term in the expansion is the magnetic quadrupole moment; a ({\P},{\T})-violating moment that appears alongside the {\P}-violating {\T}-conserving AM.

From theoretical calculations, the atomic and molecular EDMs can be linked to the hadronic and leptonic mechanisms that gave rise to them, leading to limits---and potentially values---for important fundamental physics parameters.
A summary of some of the more recent atomic and moleculer EDM measurements are presented in Table~\ref{tab:atEDM}.
Experiments have also been performed using Rb \cite{Ensberg1967a}, the excited $5p^56s\;{}^3D_2$ state of Xe \cite{Player1970}, and the YbF \cite{Hudson2011}, PbO \cite{DeMille2013}, and ThO \cite{ACME2014} molecules.
No non-zero EDM has been observed for atoms or molecules (or indeed any fundamental particle) so far; the most stringent limit $D({\rm\el{199}Hg})<3.1\E{-29}$ ($2\s$) comes from the measurements in Hg \cite{Griffith2009,Swallows2013}.

\begin{table}[t!]
\centering
\caption{Summary of the more recent atomic and molecular EDM measurements.}
\begin{tabular}{llD{.}{.}{4.16}lll} 
\toprule
\toprule
\multicolumn{2}{c}{System} & 
\multicolumn{1}{c}{EDM ($e\,\cdot\,$cm)} & 
\multicolumn{1}{c}{Year}  &
\multicolumn{2}{c}{Source}  \\
\midrule
Paramagnetic	&\el{133}Cs	&	-0.18(69)\E{-23}	&1989	&Massachusetts&\cite{Murthy1989}\\
				&\el{205}Tl		&	-0.40(43)\E{-24}	&2002	&Berkeley	&\cite{Regan2002}	\\
Diamagnetic	&\el{129}Xe 	&	0.07(33)\E{-26}		&2001	&Michigan	&\cite{Rosenberry2001}\\
				&\el{199}Hg 	&	0.049(150)\E{-28}	&2009	&Seattle		&\cite{Griffith2009}\\
Molecular		&TlF			&	-0.17(29)\E{-22}	&1991	&Yale			&\cite{Haven1991}\\
\bottomrule
\bottomrule
\end{tabular}
\label{tab:atEDM}
\end{table}

Much experimental work is currently under way that promises significant improvements in the measurements in the near future, e.g., in Xe \cite{Romalis2001a}, YbF \cite{Kara2012,Tarbutt2013}, TlF \cite{Hunter2012} and ThO \cite{Petrov2014,Loh2013,Kirilov2013,Vutha2011}.
New experiments are also in preparation designed to measure the EDM of Xe \cite{Asahi2014} and Fr \cite{Inoue2014} at CYRIC in Tohoku, Ra at KVI \cite{Santra2014} and Argonne laboratory \cite{Holt2010,Parker2012}, and Rn at TRIUMF \cite{Tardiff2013} (see also \cite{Gaffney2013}).
Very recently, the SrF molecule was successfully trapped and cooled, demonstrating the the ability of this technique, which can be applied to molecular EDM and PNC experiments \cite{Shuman2010,Barry2014}.

A proposal to use mixtures of \el{3}He and \el{129}Xe gas offers the possibility of up to four orders of magnitude improvement compared to the \el{199}Hg EDM result \cite{Jungmann2014}.
A recent proposal to use an atomic fountain experiment to measure the EDMs of alkali atoms is presented in \cite{Wundt2012}.
Experiments to search for EDMs in condensed matter systems have been proposed \cite{Ludlow2013,BudkerNatMat2010,Budker2006}, and recently performed using Eu$_{0.5}$Ba$_{0.5}$TiO$_{3}$ \cite{Eckel2012}.

\subsection{Electron EDM}\label{sec:eEDM}  %

The EDM of an electron, should it exist, can induce an EDM in an atom or molecule by interacting with the atomic field leading to the mixing of opposite-parity states (\ref{eq:atEDM}).
The magnitude of such an EDM can be expressed in the form
$  D = K d_e  $,
where $d_e$ is the electron EDM magnitude, and $K$ is an electron structure factor that comes from atomic calculations \cite{Sandars1965,Sandars1966}. 
Roughly, for heavy atoms with an external $s$ or $p_{1/2}$ electron, it can be estimated as
$K \sim 3Z^3\alpha^2R \sim 10^2$--$10^3$~\cite{Sandars1965,Flambaum1976} ($R$ is a relativistic factor). 
The factor $K$ is referred to as the electron EDM enhancement factor for obvious reasons---the EDM of an atom can be many orders of magnitude larger than the electron EDM that caused it.
In molecules, much larger enhancement 
$K \sim 10^7$--$10^{11}$
can be realized due to the mixing of the close opposite-parity rotational levels~\cite{Sushkov1978}.

Polar molecules have an exceptionally high sensitivity to an electron EDM. 
The effective electric fields inside polar molecules can exceed several GV/cm---orders of magnitude larger than any laboratory field.
The energy shift of a particle due its EDM is proportional to the external field strength, and as such an electron passing through this region would experience a significantly enhanced shift.
The effective electric field in ThO has been calculated to be 84 GV/cm, one of the largest known \cite{Skripnikov2013}. 
In early 2014, the ACME Collaboration \cite{ACME2014} exploited this technique and used ThO to place the most stringent limit on the electron EDM to date. 
They found
\begin{equation}
d_e=-2.1(37)_{\rm stat}(25)_{\rm sys}\E{-29}\;e\cdot {\rm cm},
\end{equation}
which, at the 90\% confidence level, leads to a limit of 
$|d_e|<8.7\E{-29}\;e\cdot {\rm cm}$ \cite{ACME2014}, an order of magnitude improvement over the previous best limits, which came from experiments using YbF \cite{Hudson2011} and Tl \cite{Regan2002}.
(Newer calculations suggest a slightly larger limit of $|d_e|<9.8\E{-29}\;e\cdot {\rm cm}$ \cite{Fleig2014}.)

The Tl result is the best limit on the electron EDM coming from a paramagnetic atom \cite{Regan2002}.
Using the calculated value for the enhancement factor $K=-585$ from \cite{Liu1992}, this led to the value
$  d_e= 6.9(74)\E{-28}\; e  \cdot{\rm cm}    $.
The value of $K$ for Tl is very sensitive to atomic many-body effects, but there is excellent agreement between the most complete calculations \cite{Liu1992,DzubaEDM2009,KozlovTlEDM2012}  (see also \cite{Nataraj2011} which gives a smaller result).  
The diamagnetic (closed shell) atoms are much less sensitive to the electron EDM; e.g.~the enhancement factor for Hg is $K \sim 10^{-2}$~\cite{Flambaum1985}.  
Despite this, the strong constraint on the Hg EDM~\cite{Griffith2009,Swallows2013} means the limit on electron EDM extracted from these measurements is competitive with the Tl result.

\subsection{{\P}- and {\T}-Violating Electron--Nucleon Interactions}\label{sec:PTeN}  %

The ({\P},{\T})-violating interaction between electrons with the nucleons that gives rise to atomic and molecular EDMs has the form
\begin{equation}
\hat h_{PT}^{e-N} 
= i \frac{G_F}{\sqrt{2}}\sum_N\left[
	C_N^{\rm SP}\bar{N}N\bar{e}\g_5e 
	+ C_N^{\rm PS}\bar{N}\g_5N\bar{e}e
	+ C_N^{\rm T} \bar{N}\g_5\sigma_{\mu\nu}N\bar{e}\sigma_{\mu\nu}e
\right],
\label{eq:pteN}
\end{equation}
where the summation is over nucleons, $2\sigma_{\mu\nu}=i[\g_\mu,\g_\nu]$, and $C_N^{\rm SP}$, $C_N^{\rm PS}$, and $C_N^{\rm T}$ give the strength of the scalar--pseudoscalar (SP), pseudoscalar--scalar (PS), and tensor (T) nucleon--electron interaction, respectively, see, e.g., \cite{Khriplovich1997,GingesRev2004}.

For the standard definition of the angular {\wf}s, these interactions produce real matrix elements [counter to the interaction (\ref{eq:int-pnc}), which produces imaginary matrix elements], contribute to the mixing of opposite-parity states, and hence give rise to atomic and molecular EDMs (\ref{eq:atEDM}).
Atomic calculations are required to link the induced EDM to the $C_{p,n}$ parameters;
measurements of these EDMs then lead to limits (and potentially values) for $C_{p,n}$.
Note that simple analytical formulas provide links between the matrix elements of the above PS and T interactions, meaning that, in general, calculations are only needed for one of these \cite{Flambaum1985a} (see also \cite{Khriplovich1997,GingesRev2004}).
Several such calculations have been performed; see, e.g., \cite{Dzuba2009} and references therein.
Table \ref{tab:eNlimits} lists limits on these parameters extracted from the Hg EDM measurements.

\begin{table}[b!]
\centering
\caption{({\P,\T})-violating e--N interaction limits from the \el{199}Hg experiment \cite{Griffith2009}.}
\begin{tabular}{lD{.}{.}{2.8}l} 
\toprule
\toprule
\multicolumn{1}{c}{Parameter}&\multicolumn{1}{c}{Limit}&\multicolumn{1}{c}{Calculation}\\
\midrule
$C_n^{\rm SP}$						&	6.6\E{-8}	&	\cite{Martensson1985,GingesRev2004}	\\
$C_n^{\rm PS}$						&	5.2\E{-7}	&	\cite{Dzuba2009}	\\
$C_n^{\rm T}$							&	1.9\E{-9}	&	\cite{Dzuba2009}	\\
\bottomrule
\bottomrule
\end{tabular}
\label{tab:eNlimits}
\end{table}

\subsection{Nuclear Schiff Moments}\label{sec:SM}  %

The nuclear Schiff moment (NSM) is the lowest-order ({\P},{\T})-violating term in the expansion of the nuclear potential that survives screening by the electrons.
Taking finite nuclear size into account, the effective Hamiltonian describing the interaction of electrons with the NSM can be expressed \cite{GingesSchiff2002} (see also \cite{Auerbach1996,Senkov2008})
\begin{equation}
\hat h_{\rm NSM}= - \frac{3\v{S}\cdot\v{r}}{B} \rho(\v{r}),
\label{eq:phiS}
\end{equation}
where 
$B=\int \rho(r)r^4\,\d r$,
and 
$\v{S}=S\,(\v{I}/I)$
is the NSM. 
Expressions that include a more accurate treatment of the finite-nuclear-size effects have been obtained in \cite{KozlovAScreen2012}.
The dominant mechanism that contributes to the NSM is believed to be the ({\P,\T})-violating nucleon--nucleon interaction. 
Though EDMs of the protons and neutrons don't directly contribute to the EDM of a neutral atom (due to the screening), they can in fact induce an EDM via their contribution to the NSM, meaning that limits for $p$ and $n$ EDMs can be obtained from NSM measurements.

The interaction (\ref{eq:phiS}) leads to mixing of opposite-parity states, and thus contributes to the atomic and molecular EDM (\ref{eq:atEDM}).
In order to extract a value for the NSM from EDM measurements, atomic calculations are required.
Nuclear calculations are required to link the NSM to the parameters of the inter-nucleon ({\P,\T})-violating interaction \cite{Sushkov1984}.

Atomic calculations for many atoms of experimental interest have been performed \cite{Flambaum1985b} (see also, e.g.,  \cite{FlambaumNPA1986,DzubaEDM2002,Latha2009,Dzuba2009,Singh2014}).
Calculations have also been performed for molecules (see, e.g., \cite{Kudashov2013}) and solid-state systems \cite{Ludlow2013}.
There are several recent nuclear many-body calculations, though the agreement is not ideal; see, e.g., \cite{Engel2013,Dekens2014} and references therein.
From the Hg measurements \cite{Griffith2009}, a limit on the constant $S$ can be found:
\begin{equation}
  S(^{199}{\rm Hg}) < 1.2\E{-12} \;e\cdot{\rm fm}^3.
\label{eq:x}
\end{equation}
For consistency, we follow \cite{Swallows2013}, and use both the atomic and nuclear calculations from \cite{Dzuba2009}.
From the nuclear calculations \cite{Dmitriev2003} (see also \cite{FlambaumNPA1986,DzubaEDM2002}), the NSM can be expressed in terms of the proton and neutron EDMs: $S(\el{199}Hg)=1.9d_n+0.2d_p$ (see \cite{Dmitriev2003} for a discussion on the uncertainty).
From this, the limits
$d_p < 8.6\E{-25} e \ {\rm cm}$
and
$d_n  < 6.3\E{-26} e \ {\rm cm}$
can be extracted \cite{Swallows2013}.
Nuclear and quantum chromodynamics (QCD) \cite{Crewther1979} calculations can also link the induced EDM to the observable strong {\C\P}-violation parameter $\bar\theta_{\rm QCD}$ \cite{Haxton1983a}, and the difference between the up- and down-quark chromo-EDMs \cite{Pospelov2002}. 
From the calculations of \cite{Crewther1979,Pospelov1999,Pospelov2002}, the limits
$\bar\theta_{\rm QCD}<5.3\E{-10}$
and
$|\tilde d_u-\tilde d_d|<6\E{-27} \, {\rm cm}$
are placed (see also \cite{Swallows2013}).

It was pointed out in \cite{Auerbach1996}, that the NSM may be strongly enhanced in nuclei
that possess octupole deformation.
Nuclear deformation creates an intrinsic Schiff moment, $S_{\rm int}$, in the rotating (frozen) nuclear reference  frame, which is zero in the lab frame and does not violate {\P}- or {\T}-invariance.
Alone, this can not lead to mixing of opposite parity states.
However, the ({\P,\T})-violating internuclear interaction mixes opposite parity levels of the doublet and polarises nuclear axis along nuclear spin, allowing a nonzero NSM to appear on the lab frame.
The small energy intervals between these doublet states leads to an enhancement of several orders of magnitude of the NSM in the laboratory frame.

\subsection{Nuclear Magnetic Quadrupole Moments}\label{sec:MQM}

The lowest order ({\P},{\T})-violating magnetic moment of the nucleus is the magnetic
quadrupole moment (MQM).
Borne of the same {\C\P}-odd nuclear forces as the NSM, the MQM can lead to permanent EDMs of atoms and molecules by mixing electronic states of opposite parity.

The magnitude of the NSM can be roughly estimated as $S \sim r_N^2 d_N$, where $r_N\sim 1$ fm  is the nuclear radius. 
The smallness of $r_N$ means that the atomic EDM produced by the NSM is much smaller than $d_N$.
In \cite{Sushkov1984}, it was demonstrated the MQM produces a larger EDM than the NSM in paramagnetic atoms and molecules (see also \cite{GingesRev2004,Khriplovich1997}).
Also, for deformed nuclei, the MQM has a collective nature and is strongly enhanced \cite{Flambaum1994} (see also \cite{FlambaumMQM2014}).
In contrast to the SM, ordinary quadrupole deformation (which exists in 50\% of nuclei, and in all nuclei of experimental interest) is sufficient to allow collective enhancement of the MQM.
Despite the significant enhancement, it has been a large challenge to design experiments that are sensitive to the ({\P,\T})-odd hadronic physics manifested in the form of MQMs.
However, recent advances mean that the significant advantages of MQMs can be now exploited in experiments.
In particular, it has become possible to perform EDM measurements using molecules in paramagnetic ${}^3\Delta_1$  states \cite{Meyer2006}.
Such systems have an $\Omega$-doublet substructure, allowing for full polarization in modest external electric fields  \cite{Sushkov1978}. 
Intense, slow molecular beams \cite{Hutzler2011,Barry2011}  and techniques for spin-precession measurements both on such beams \cite{Kirilov2013} and on trapped molecular ions \cite{Loh2013} have been developed, and implemented \cite{ACME2014}.

In \cite{FlambaumMQM2014}, the possibility of using ${}^3\Delta_1$  molecular states to search for ({\P,\T})-odd interactions in the hadron sector was considered in detail.
This approach takes advantage of the dramatically enhanced energy shifts associated with the strong electric polarization of molecules, as well as the enhanced effects of the MQM, especially in deformed nuclei.
This should greatly increase the sensitivity compared to the $^{199}$Hg atomic experiment \cite{Griffith2009}, where the limit on the NSM now places the strongest limits on most underlying {\C\P}-odd effects.

As in the case of the NSM, atomic and molecular calculations are required to link the induced EDM to the MQM, and nuclear calculations are required to link the MQM to the underlying {\C\P}-odd nuclear interactions.
Due to differences in the nuclear structure, it appears that the interpretation of the MQM may actually be more reliable than that of the NSM \cite{FlambaumMQM2014}, presenting a further advantage of using the MQM to probe low-energy hadronic physics.

The NSM, an electric moment, is affected by screening from the atomic electrons.
In practical calculations this presents a major source of instability \cite{Sushkov1984}, making them particularly sensitive to finite-nuclear-size \cite{GingesSchiff2002,KozlovAScreen2012} and many-body \cite{Flambaum1985b,FlambaumNPA1986,Ban2010,Dmitriev2005} effects.
For systems such as \el{199}Hg that have a valence neutron, the NSM is generated mainly through polarization of the nuclear core  \cite{Flambaum1985b} (see also \cite{FlambaumNPA1986}); the contribution from the valence neutron is zero.
This acts to significantly suppress the effects, and also greatly increases the instability of the calculations \cite{Dmitriev2005a,Ban2010}.
For the MQM, however, there is no such screening, and the valence nucleon
gives the main contribution \cite{Dmitriev1996}.

The effect of MQMs for many heavy molecules was first calculated in \cite{Sushkov1984}.
More recent calculations of the necessary molecular structure were performed in \cite{Kozlov1995}  for BaF, YbF, and HgF, and in \cite{Skripnikov2014} for ThO.
In \cite{FlambaumMQM2014}, a detailed study was performed of the MQM effects in many diatomic molecules. 
It was found that the sensitivity to nuclear ({\P,\T})-violating effects is high in
paramagnetic molecules containing deformed nuclei. 
The authors conclude that if measurements of EDM-like frequency shifts can be made with a sensitivity of around an order of magnitude better than in the recent electron EDM experiment using ThO molecules \cite{ACME2014}, then limits on several underlying parameters of hadronic {\C\P}-violation can be improved.

\section{DARK-MATTER AXION DETECTION}\label{sec:cosmic}

Among the most important unanswered questions in fundamental physics are the strong {\C\P} problem---the ``unnatural'' smallness of the $\theta_{\rm QCD}$ parameter in the QCD Lagrangian that quantifies the amount of {\C\P}-violation \cite{Weinberg1976}---and the question of dark matter and dark energy, see e.g., \cite{Bertone2005,Riess1998}.
One elegant solution to the strong {\C\P} problem invokes the introduction of a pseudoscalar particle known as the axion~\cite{Peccei1977a}.
It has been noted that the axion may also be a promising cold dark matter (CDM) candidate. 
Thus axions, if detected, could resolve both the CDM and strong {\C\P} problems.

The prospect of using atomic systems to search for axions has been considered broadly in the literature; for a recent review, we direct the reader to \cite{StadnikRev2014}.
Here, we discuss the recent proposal to use the PNC amplitudes and EDMs that are induced in atomic systems to search for evidence of axions and other cosmic fields \cite{Stadnik2014,\RobertsCosmic}.
Such searches would be complementary to proposals to use axion-induced {\P}-conserving $M1$ transitions \cite{Sikivie2014}.

The interaction of a pseudoscalar cosmic field with electrons can be described by the Lagrangian density
\begin{equation}
\mathcal{L}^{\rm PS} =
 i \zeta m_e\, \phi  \,  \bar e \gamma_5  e 
- \eta (\partial_\mu\phi)\, \bar e  \gamma^\mu\gamma_5 e
\label{eq:lps},
\end{equation} %
where $\zeta$ and $\eta$ are dimensionless constants quantifying the interaction strength (including the field amplitudes) of fermions with the field via a direct  and derivative-type coupling, respectively, and
$m_e$ is the electron mass.
In the case of axions, $\phi=\phi(\v{r},t)$ represents the dynamic axion field that obeys the Klein-Gordon equation, and can be expressed $\phi(\v{r},t) = \cos(\omega_a t)$ for a particular choice of phase.

The PNC amplitudes and EDMs induced by these interactions can be expressed
\begin{align}
&E_{\rm PNC}^{\rm PS}(\zeta) = \frac{\zeta\hbar\omega_a}{2}\sin(\omega_a t) K_{\rm PNC}	\label{eq:pncg0g5}\\
&E_{\rm PNC}^{\rm PS}(\eta) = {\eta \hbar\omega_a}\sin(\omega_a t) K_{\rm PNC} \label{eq:pncg5}\\
&d_{\rm EDM}^{\rm PS}(\zeta) =  {-i\zeta \hbar^2\omega_a^2} \cos(\omega_a t) K_{\rm EDM} \label{eq:edmg0g5}\\
&d_{\rm EDM}^{\rm PS}(\eta) = {-2i\eta  \hbar^2\omega_a^2 \cos(\omega_a t)} K_{\rm EDM}, \label{eq:edmg5}
\end{align}
where $\hbar\omega_a$ is the energy of the field particle (e.g.~the axion)  \cite{\RobertsCosmic} (see also \cite{Stadnik2014}).
In the above equations $K_{\rm PNC}$ and $K_{\rm EDM}$ are atomic structure factors, which were calculated in \cite{\RobertsCosmic}.
Such fields can also create oscillating AMs, NSMs, and MQMs via interaction with the nucleus \cite{Stadnik2014,\RobertsCosmic}.

To detect these dynamic effects, an experiment designed to  measure small oscillations in the PNC amplitude or atomic EDM is needed. 
The frequency and amplitude of these oscillations would enable one to extract values for the relevant field parameters. 
For example, a determination of the frequency would provide the mass of the particle, and the amplitude of the oscillations would lead to a determination of the constants $\eta$, $\zeta$.

The high sensitivity of atomic EDM experiments makes them promising for the study of the oscillating effects considered here.
Further enhancement in the sensitivity of the EDM measurements can be obtained by tuning the experiment to a specific frequency; see, e.g., \cite{Graham2013,Budker2014,Stadnik2014}. 
It is possible that the oscillating EDMs considered here could be measured with a higher sensitivity than static EDMs, since the dominating source of of systematic uncertainty in these measurements stems from the reversal of the electric field---an operation that is redundant when considering an oscillating effect \cite{Budker2014}.
Axions with masses $m_a=10^{-5}$ eV/$c^2$ or $10^{-9}$ eV/$c^2$
would lead to oscillations with frequencies on the order of GHz or MHz, respectively. 
The coherence time may be estimated from 
$\Delta \omega_a/ \omega_a\sim(\tfrac{1}{2}m_av^2/m_ac^2)\sim(v^2/c^2)$, 
where a virial velocity of $v \sim 10^{-3} c$ would be typical in our local Galactic neighbourhood, and $\omega_a\approx m_ac^2/\hbar$~\cite{Graham2011}.
It is also important to note that the effects considered here are linear in the small parameter ($\eta$,$\zeta$) that quantifies the interaction between dark matter and ordinary matter particles.
Most other current dark matter axion searches rely on effects that are proportional to quadratic and higher powers of this parameter.

As well as pseudoscalar axion fields, other cosmic fields can also induce {\P}-violating effects in atoms.
Such fields have been considered in, e.g., Lorentz-invariance-violating standard-model extensions~\cite{Colladay1997}. 
In \cite{\RobertsCosmic}, the {\P}-violating interaction of the temporal component of a pseudovector cosmic field with fermions, which has the form
\begin{equation}
\mathcal{L}^{\rm PV} = b_0\bar\psi \gamma^0\gamma^5\psi   ,
\end{equation}
was considered.
By combining calculations of the PNC effects induced by such a field with existing experiments, limits were obtained on the parameter $b_0$.
The most stringent limits for the interaction with
 electrons, $|b_0^e| < 7 \times 10^{-15}$ GeV \cite{\RobertsCosmic}, comes from the experiments in Dy \cite{Nguyen1997},
and the best limits for the interaction with protons, 
$|b_0^p| < 4 \times 10^{-8}$ GeV \cite{\RobertsCosmic,StadnikNMB2014}, 
and  neutrons, 
$|b_0^n| < 2 \times 10^{-7}$ GeV \cite{\RobertsCosmic,StadnikNMB2014},
 come from the experiments in  Cs \cite{Wieman1997}.
These limits on the temporal components, $b_0$,
 which are derived  from {\P}-violating effects, 
are complementary to existing limits on the  spatial components, $\v{b}$, derived from {\P}-conserving effects due to the interaction of static cosmic fields with electrons, protons and neutrons, of
$ 1.3 \times 10^{-31}$ GeV~\cite{Heckel2006}, 
$1.6 \times 10^{-33} $ GeV~\cite{StadnikNMB2014}
and 
$8.4 \times 10^{-34}$ GeV~\cite{Allmendinger2014},
 respectively.
For further details and a brief history on recent developments in these limits, we refer the reader to \cite{Kostelecky2011a}.

The detection of axions and axion-like particles in the form of topological-defect dark matter has also been suggested through their interactions with fermion spins \cite{Pospelov2013}, and the induction of transient EDMs \cite{StadnikDefects2014}. 
Such effects can be searched for using a global network of magnetometers \cite{Pospelov2013} or atomic clocks \cite{DereviankoDM2014}.

\paragraph{Acknowledgements}
We would like to thank 
Y.~Stadnik 
and 
B.~Lackenby
for reading the manuscript and for their useful suggestions,
as well as 
A.~Kozlov
and J.~Berengut
for many helpful discussions.
This work was supported by the Australian Research Council.
V.~Flambaum would also like to acknowledge the Humboldt foundation for support in the form of the Humboldt Award, and the MBN Research Center, where part of this work was conducted, for hospitality.
The authors are not aware of any affiliations, memberships, funding, or financial holdings that might be perceived as affecting the objectivity of this review.

\footnotesize 	
\bibliography{../AllReferences/library}

\end{document}